\documentclass[prd,twocolumn,superscriptaddress,showpacs,nofootinbib,preprintnumbers]{revtex4}

\usepackage{amsmath}
\usepackage{amsfonts}
\usepackage{graphicx}
\usepackage{dcolumn}
\usepackage{hyperref}

\newcommand{\be}{\begin{equation}}
\newcommand{\ee}{\end{equation}}
\newcommand{\bea}{\begin{eqnarray}}
\newcommand{\eea}{\end{eqnarray}}
\newcommand{\beaa}{\begin{eqnarray*}}
\newcommand{\eeaa}{\end{eqnarray*}}

\newcommand{\nn}{\nonumber \\}

\newcommand{\e}{{\rm e}}

\begin{document}

\tolerance=5000

\title{The new form of the equation of state for dark energy fluid
and accelerating universe}
\author{Shin'ichi Nojiri}
\email{nojiri@phys.nagoya-u.ac.jp}
\affiliation{Department of Physics, Nagoya University, Nagoya 464-8602. Japan}
\author{Sergei D. Odintsov\footnote{also at Lab. Fundam. Study, Tomsk State
Pedagogical University, Tomsk}}
\email{odintsov@ieec.uab.es}
\affiliation{Instituci\`{o} Catalana de Recerca i Estudis Avan\c{c}ats (ICREA)
and Institut de Ciencies de l'Espai (IEEC-CSIC),
Campus UAB, Facultat de Ciencies, Torre C5-Par-2a pl, E-08193 Bellaterra
(Barcelona), Spain}

\begin{abstract}

We suggest to generalize the dark energy equation of state (EoS) by
introduction
the relaxation equation for pressure which is equivalent to consideration
of the inhomogeneous EoS cosmic fluid which often appears as the
effective model from strings/brane-worlds. As another, more wide
generalization we discuss the
 inhomogeneous EoS which contains derivatives of pressure.
For several explicit examples motivated by the analogy with classical
mechanics the accelerating FRW cosmology is constructed.
It turns out to be the asymptotically de Sitter or oscillating universe
with possible  transition from deceleration to acceleration phase.
The coupling of dark energy with matter in accelerating FRW universe is
considered, it is shown to be
consistent with constrained (or inhomogeneous) EoS.

\end{abstract}


\maketitle

\section{Introduction}

The number of attempts is aimed to the resolution of dark energy problem
(for recent review, see \cite{rev1,rev2,rev3}) which is considered  as the
most fundamental one in modern cosmology.
Among the different descriptions of  late-time universe the easiest one
is phenomenological approach where it is assumed that universe is filled
with mysterious cosmic fluid of some sort. One can mention imperfect
fluids \cite{capozziello},
general equation of state (EoS) fluid where pressure is some (power law)
function of energy-density \cite{gen},
fluids with inhomogeneous equation of state \cite{inh},
where EoS with time-dependent bulk viscosity is the particular case \cite{brevik},
coupled fluids \cite{coupled1,coupled2}, etc.
The EoS fluid description may be even equivalent to modified gravity
approach as is shown in \cite{modified}.

As it has been recently discussed in \cite{amendola} it is not easy
to construct the dark energy model which describes the universe
acceleration and  on the same time keep untouched the radiation/matter
dominated epochs with subsequent transition from deceleration to
acceleration. In order to minimize the dark energy effect at intermediate
epoch one may speculate about sudden appearence of dark energy around the
deceleration-acceleration transition point. In other words, one may
suppose that EoS of DE fluid is of the form $p=\theta (t-t_d)w\rho$
where $t_d$ is transition time and $w$ is DE EoS parameter.
Before transition point, DE plays a role of usual dust which changes EoS
by some
unknown scenario.
In the similar way, one can generalize other cosmic fluids with more
complicated EoS. This introduces the idea of structure/form changing EoS
in different epochs. The simplest example of such cosmic fluid is
oscillating dark energy \cite{oscil,osc}. Finally, the reason why
cosmic fluid still escapes of direct observations
could be that it has completely unexpected properties, for instance, in
EoS picture.

In the present letter the new form of dark energy EoS is considered.
As the first step, we introduce the relaxation equation for pressure (the
analog of energy conservation law for energy-density). It is then
shown that such constrained EoS is equivalent to usual but inhomogeneous
EoS
which is known to be the effective description for brane-worlds or
modified gravity \cite{inh}.
The generalized inhomogeneous EoS which contains time
derivatives of pressure is introduced.
The number of examples for such EoS cosmic fluids is presented
and the corresponding FRW cosmologies are described.
It is shown that cosmic speed-up in the examples under consideration
corresponds to the
asymptotically de Sitter  or the oscillating universe where
accelerating/decelerating epochs repeat with possibility
 to cross the phantom barrier or to make transition
from deceleration to acceleration. In all cases, dark energy
EoS parameter is close to $-1$, being within the observational bounds to
it.
It is demonstrated that  the inclusion of matter may be consistent
 with constrained EoS.

\section{Constrained equation of state}

Let us discuss the possible modification of the equation of state
in such a way that it would change its structure/form during the
universe evolution.
Consider the balance equation for the energy (energy conservation law)
\be
\label{B1}
\dot{\rho} + 3 H (\rho + P) = 0 \,.
\ee
It can be represented as a relaxation equation
\be
\label{B2}
\dot{\Psi} = - \frac{1}{\tau} (\Psi - \Psi_0) \,,
\ee
where $\Psi$ coincides with $\rho$, relaxation time $\tau$ is
$\tau = \frac{1}{3H}$ and stationary (or equilibrium) value of $\rho$ is
$\rho_0 \equiv \Psi_0 = - P$.
To formulate consistently this equation we need, as usual, the equation of
state (EoS). The standard barotropic EoS is $P = P(\rho)$, providing the
equation for $\rho$ only:
\be
\label{B3}
\frac{1}{3H} \dot{\rho} = - (\rho + P(\rho)) \,.
\ee

Let us conjecture now that cosmological fluid is described by different EoS at
different epochs. In other words, to describe the transition from one epoch in
cosmological evolution
to another we try to introduce the transition from one EoS to another,
or in simplest form to modify EoS to permit the presence of pressure
derivatives.

The simplest way is to introduce the relaxation equation for pressure
\be
\label{B4}
\tau \dot{P} + P = f(\rho, a(t)) \,.
\ee
When $\tau = 0$ and $f(\rho, a(t)) = P(\rho)$ we recover the
standard EoS. Such an equation may be considered as some (dynamical) constraint
to usual EoS.
Of course, the physical sense of such equation (unlike to
energy conservation law) is not clear at the moment although some explanations
are given in the next section.
In daily life, however, there could occur similar phenomena where the time
change of
the presure depends on the density. For example, consider the water. There
is a
pressure in the steam, which is the gas of water. When the density increases,
the molecules of water make drops of water, like
fog. The pressure of the drops could be neglected. At high density,
the total
pressure could decrease. The equation (\ref{B4}) seems to express
such a process.
Then if dark energy consists of particles or some objects with internal
structure,
there may occur the phase transition like that between steam and water drop.
At the point of phase transition, since the system becomes unstable, the
pressure
may be governed by a equation like (\ref{B4}).

We prefer to measure the relaxation time $\tau$ in terms of Hubble
function $H$, i.e., consider $\tau H = \xi = const$. In this case it is
convenient to use a new variable $x \equiv \frac{a(t)}{a(t_0)}$.
In terms of $x$ the expression $\tau \dot{P}$ simplifies as
\be
\label{B5}
\tau \dot{P} = \xi x \frac{d P}{dx}
\ee
and one obtains the pair of relaxation type equations for $\rho$ and $P$
\bea
\label{B6}
&& \frac{1}{3} x \frac{d \rho}{dx} + \rho = - P   \,, \\
\label{B7}
&& \xi x \frac{d P}{dx} + P = f(\rho, x) \,.
\eea
Extracting $P$ from the first equation and inserting it to the second one
we obtain the second order, master equation  for the energy density
$\rho$
\be
\label{B8}
x^2 \frac{d^2 \rho}{dx^2} + x \frac{d \rho}{dx} \left(4 + \frac{1}{\xi} \right)
+ \frac{3}{\xi} [\rho + f(\rho,x)] = 0
\ee
This is new, dynamical equation to energy-density which is compatible with
energy conservation law.

As the explicit example,
let the function $f$ be of the form (of course, more complicated choices
may be considered)
\be
\label{B9}
f(\rho,x) = - \rho + \gamma(x) (\rho -\rho_c) \,,
\ee
where $\rho_c = \mbox{const}$ is some critical value of the energy density,
and
$\gamma(x) = \gamma_0 + \alpha x^m$.
When $\rho_c = 0$ and $\alpha =0$ we recover the standard linear EoS
\be
\label{B10}
P=(\gamma_0 -1)\rho \,.
\ee

The equation for $\rho$ can be
reduced to the Bessel equation and the solution is of the form
\be
\label{B11}
\rho = \rho_c + x^{\sigma} \left[C_1 J_{\nu}(\hat a x^{\lambda}) +
C_2 J_{-\nu}(\hat a x^{\lambda})\right] \,.
\ee
Here
\be
\label{B11b}
\sigma = - 2 - \frac{m}{2} - \frac{1}{\xi}\ ,\quad \lambda=\frac{m}{2}\ ,\quad
\hat a=\alpha^{m/2}\ .
\ee
Using (\ref{B6}), we find
\bea
\label{B11c}
P&=& - \rho_c \nn
&& - \left(\frac{\sigma}{3} + 1\right)x^{\sigma} \left[C_1 J_{\nu}(\hat a
x^{\lambda}) +
C_2 J_{-\nu}(\hat a x^{\lambda})\right] \nn
&& - \frac{\lambda \hat a}{6}x^{\sigma+\lambda}
\left[C_1 \left(J_{\nu - 1}(\hat a x^{\lambda}) - J_{\nu + 1}(\hat a
x^{\lambda}) \right) \right. \nn
&& \left. + C_2 \left(J_{- \nu - 1}(\hat a x^{\lambda}) - J_{-\nu + 1}(\hat a
x^{\lambda})\right)\right]\ .
\eea
When $x$ is large, $\rho$ behaves as an oscillating function
\bea
\label{B11d}
\rho &\sim& \rho_c + \sqrt{\frac{2}{\pi \hat a}} x^{\sigma - \lambda/2}
\left\{ C_1 \cos \left( \hat a x^\lambda - \frac{2\nu + 1}{4}\pi\right) \right.
\nn
&& \left. + C_2 \cos \left( \hat a x^\lambda - \frac{-2\nu + 1}{4}\pi\right)
\right\}\ .
\eea
Since $\sigma - \lambda/2 = -2 - (3/4)m - 1/\xi$, if we naturally assume that
$m$ and $\xi$
should be positive, the second term damps with oscillation. Then $\rho$ goes to
a constant $\rho\to \rho_c$.
On the other hand, for large $x$, $P$ behaves as
\bea
\label{B11e}
P &\sim& - \rho_c \nn
&& - \frac{\lambda \hat a}{3} \sqrt{\frac{2}{\pi \hat a}} x^{\sigma +
\lambda/2}
\left\{ C_1 \cos \left( \hat a x^\lambda - \frac{2\nu - 1}{4}\pi\right) \right.
\nn
&& \left. + C_2 \cos \left( \hat a x^\lambda + \frac{-2\nu + 1}{4}\pi\right)
\right\}\ .
\eea
Since $\sigma + \lambda/2 = -2 - m/4 - 1/\xi$, when $m$ and $\xi$ are positive,
the second term
damps with oscillation again and $P$ goes to a constant $P\to - \rho_c$.
Then the effective EoS parameter $w\equiv P/\rho$ goes to $-1$, which
corresponds to a cosmological
constant.

The Bessel function is quasi-oscillating and we obtain an infinite number
of epochs, in which $\rho$, $P$, $H$ and $a$ are also quasi-oscillating. In
other words we have an infinite number of points in which the deceleration
replaces the acceleration and vice-versa.

The presence of $\rho_c$ can guarantee that $\rho$ is positive, thus,
$H^2$ is
also positive. Nevertheless, $P$ can change its sign, and this phenomenon
can mimic the dark energy effect.

When $\rho_c = 0$, $\alpha =0$ the equation becomes of the Euler type, and
the solution is also very simple.

\section{The relation with standard equation of state.}

We now consider the relation with the standard EoS.
Let us start with the scale factor dependent EoS:
\be
\label{B12}
P=g\left(\rho,a\right)\ .
\ee
Then we have
\bea
\label{B13}
&& \e^{-t/\tau}\frac{d}{dt}\left(\e^{t/\tau} P \right) \nn
&& = \frac{1}{\tau}P + \dot P \nn
&& = \frac{1}{\tau} g\left(\rho,a\right) + \frac{\partial
g\left(\rho,a\right)}{\partial \rho}\dot\rho
+ \frac{\partial g\left(\rho,a\right)}{\partial a} a H\ .
\eea
By using the conservation law (\ref{B1}), one can rewrite (\ref{B13}) in a
form
similar to (\ref{B4}):
\bea
\label{B14}
&& \tau \dot P + P \nn
&& = g\left(\rho,a\right) + \tau H\left( -3 \frac{\partial
g\left(\rho,a\right)}{\partial \rho}
\left(\rho + g\left(\rho,a\right)\right) \right. \nn
&& \left. + \frac{\partial g\left(\rho,a\right)}{\partial a} a\right)\ .
\eea
When  other contributions to the energy density
can be neglected,  the first FRW equation looks as
\be
\label{B15}
\frac{3}{\kappa^2}H^2=\rho\ .
\ee
Then Eq.(\ref{B14}) can be rewritten as
\bea
\label{B16}
&& \tau \dot P + P \nn
&& = g\left(\rho,a\right) + \tau \kappa \sqrt{\frac{\rho}{3}}\left( -3
\frac{\partial g\left(\rho,a\right)}{\partial \rho}
\left(\rho + g\left(\rho,a\right)\right) \right. \nn
&& \left. + \frac{\partial g\left(\rho,a\right)} a\right)\ .
\eea
By comparing (\ref{B16}) with (\ref{B4}), we may identify
\bea
\label{B17}
&& f(\rho,a) \nn
&& =g\left(\rho,a\right) + \tau \kappa \sqrt{\frac{\rho}{3}}\left( -3
\frac{\partial g\left(\rho,a\right)}{\partial \rho}
\left(\rho + g\left(\rho,a\right)\right) \right. \nn
&& \left. + \frac{\partial g\left(\rho,a\right)} a\right)\ .
\eea
This shows the relation between standard (generally speaking,
inhomogeneous EoS\cite{inh}) and relaxation equation for pressure.

\section{Generalized inhomogeneous equation of state}

As it was indicated above, there is a possibility that the EoS contains
$\dot P$
or even higher time derivatives of pressure.
More generally, the EoS could
depend on $H$ or $\dot H$ (inhomogeneous EoS \cite{inh}) like
\be
\label{B19}
U(\rho,P,\dot P, H, \dot H)=0\ .
\ee
Note that many effective dark energy models like brane-worlds, modified
gravity and string compactifications have such a form ( for very
recent example compatible with observational data, see \cite{roy} and
references therein).
 As particular example, one may consider
\be
\label{B20}
U(\rho,P,\dot P, H, \dot H)=
\dot P + \left(\frac{\dot H}{H} - 3H\right)\left(\rho + P\right) +
\frac{W(\rho)}{3H}\ .
\ee
Here $W(\rho)$ is a proper function of the energy density $\rho$. Using
the energy conservation law
(\ref{B1}), one gets
\be
\label{B21}
\ddot \rho = W(\rho)\ .
\ee
If  $\rho$ is regarded as a coordinate, Eq.(\ref{B21}) has a form of
Newtonian
equation of motion of the
classical particle with the ``force'' $W$.
For example, if $W$ is a constant $W=w_0$, we find $\rho$ behaves as a
coordinate of the massive particle in the
uniform gravity:
\be
\label{B22}
\rho=\frac{w_0}{2}\left(t - t_0\right)^2 + c_0\ .
\ee
Here $t_0$ and $c_0$ are constants of the integration.
As an another example, we may consider the case of the harmonic oscillator:
\be
\label{B23}
W(\rho) = - \omega^2 \left(\rho - \rho_0\right)\ .
\ee
Then  an oscillating energy density follows \cite{oscil,osc}:
\be
\label{B24}
\rho=\rho_0 + A\sin \left(\omega t + \alpha \right)\ .
\ee
If  other contributions to the energy density may be neglected, by using
the first
FRW equation (\ref{B15}),
we find the behavior of the Hubble rate, for (\ref{B22}),
\be
\label{B25}
H=\frac{\kappa}{\sqrt{3}}\sqrt{\frac{w_0}{2}\left(t - t_0\right)^2 + c_0}\ .
\ee
As $\dot H<0$ when $t<t_0$, and $\dot H>0$ when $t>t_0$, there is a transition
from non-phantom era to
phantom one at $t=t_0$.
For (\ref{B24}), we have oscillating $H$:
\be
\label{B26}
H=\frac{\kappa}{\sqrt{3}}\sqrt{\rho_0 + A\sin \left(\omega t + \alpha \right)}\
.
\ee
When we neglect the other contributions to the energy density and pressure, we
also have
\be
\label{B29}
-\frac{2}{\kappa^2}\dot H = \rho + p\ .
\ee
Combining (\ref{B29}) with (\ref{B15}), one may define the effective EoS
parameter $w_{\rm eff}$ by
\be
\label{Bl1}
w_{\rm eff}\equiv -1 - \frac{2\dot H}{3H^2}\ .
\ee
Hence, for (\ref{B25})
\be
\label{Bl2}
w_{\rm eff}= - 1 - \frac{t-t_0}{\sqrt{3}\kappa \left(\frac{w_0}{2}
\left(t - t_0\right)^2 + c_0\right)^{3/2}}\ ,
\ee
which surely crosses $w_{\rm eff}=-1$ when $t=t_0$.
On the other hand, for (\ref{B26}), one gets
\be
\label{Bl3}
w_{\rm eff}= - 1 - \frac{2A\omega \cos \left(\omega t +
\alpha\right)}{\sqrt{3}\kappa
\left(\rho_0 + A\sin \left(\omega t + \alpha \right)\right)^{3/2}}\ ,
\ee
which  oscillates around $w_{\rm eff}=-1$ as in \cite{osc}.

As an another example, we consider the EoS
\be
\label{B26b}
\dot P - 3H\left(\rho + P\right)=U(H)\ .
\ee
Here $U(H)$ is a proper function of the Hubble rate $H$.
Then by using (\ref{B1}), one arrives at
\be
\label{B27}
\dot\rho + \dot P=U(H)\ .
\ee
In a simplest case, $U(H)=0$, it follows
\be
\label{B28}
\rho + P = c\quad (c:\mbox{constant})\ .
\ee
When  the other contributions to the energy density and pressure are
neglected,
because of (\ref{B29}), we find $\dot H$ is constant and
\be
\label{B30}
H=-\frac{\kappa^2c}{2}t\ .
\ee
As an another case, we may consider
\be
\label{B31}
U(H)=\frac{2\omega^2}{\kappa^2}H\ .
\ee
Here $\omega$ is a constant. Then combining (\ref{B27}), (\ref{B29}), and
(\ref{B31}), we find
\be
\label{B32}
\ddot H = - \omega^2 H\ ,
\ee
which is the equation typical for the harmonic oscillator in classical
mechanics.
Hence, the oscillating Hubble rate is obtained
\be
\label{B33}
H=H_0 \sin \left(\omega t + \alpha\right)\ .
\ee
Here $H_0$ and $\alpha$ are constants of the integration.
Thus, we demonstrated that inhomogeneous generalized EoS (linear in the
 pressure derivative)
leads to the interesting accelerating (often oscillating) late-time
universe.

\section{The equation of state quadratic on the pressure derivative}

In this section, as an immediate generalization, the case that the EoS is not linear
on $\dot P$ but quadratic is considered.

Let the equation to pressure and its derivatives looks like an energy in
the classical mechanics:
\be
\label{B34}
E=\frac{1}{2}{\dot P}^2 + V(P)\ .
\ee
Here $E$ is a  constant but as it is an analogue of the energy,
it is denoted as $E$. We should note that $E$ does not correspond to
real energy in universe. This may be also considered as implicit form of
EoS.

First example is
\be
\label{B35}
V(P)=\tilde a P\ ,
\ee
with a constant $\tilde a$. Then by the analogy with the classical
mechanics,
we find
\bea
\label{B36}
P&=&-\frac{1}{2}\tilde at^2 + v_0 t + p_0\ ,\nn
E&=& \frac{1}{2}v_0^2 + \tilde a p_0\ .
\eea
Here $v_0$ and $p_0$ are constants.

In case that other contributions to the total energy density are large, as in the early
universe, the Hubble rate $H$ could not be so rapidly changed. Then we may assume that
the Hubble rate $H$ could be almost constant $H=H_0$.
Using (\ref{B1}), one obtains
\bea
\label{B37}
\rho &=& \rho_0 \e^{-3H_0 t} \nn
&& - \frac{\tilde a}{2}\left(\frac{2}{27H_0^3} - \frac{2t}{9H_0^2} +
\frac{t^2}{3H_0}\right) \nn
&& - v_0\left(- \frac{1}{9H_0^2} + \frac{t}{3H_0}\right) + \frac{p_0}{3H_0}\ .
\eea
Here $\rho_0$ is a constant. The explicit form of (inhomogeneous) EoS
may be found combining two above equations.

On the other hand, we may also consider the case that the other contributions to the
energy density and pressure are neglected as in late-time or future unverse.
Then deleting $\rho$ from (\ref{B1}) and (\ref{B15}), we have
\be
\label{BBB1}
\dot H + \frac{3}{2}H^2 + \frac{\kappa^2}{2}P=0\ .
\ee
For (\ref{B36}), Eq.(\ref{BBB1}) admits the solution
\be
\label{BBB2}
H=h_0 t + h_1\ ,
\ee
when
\bea
\label{BBB3}
&& \tilde a= 3h_0\ ,\quad v_0=-3h_0 h_1\ , \nn
&& p_0 = - h_0 - h_1^2\ .
\eea
For (\ref{BBB2}), the effective EoS parameter $w_{\rm eff}$ defined by
(\ref{Bl1}) has the following form:
\be
\label{BBB4}
w_{\rm eff}=-1 - \frac{2h_0}{3\left(h_0 t + h_1\right)^2}\ ,
\ee
which goes to $-1$ when $t$ goes to infinity.
Hence, the emerging universe seems to be the asymptotically de Sitter one.

Second example is
\be
\label{B38}
V(P)=\frac{1}{2}\omega^2 P^2\ .
\ee
Then we have
\bea
\label{B39}
P&=&A\sin\left(\omega t + \alpha\right) \nn
E&=& A^2 \omega^2\ .
\eea
where $A$ and $\alpha$ are constants. Then in the case
that other contribution to the total energy density is large, as in the early
universe, the Hubble rate $H$ could be almost constant $H=H_0$,
we find
\bea
\label{B40}
\rho &=& \rho_0 \e^{-3H_0 t} \nn
&& - \frac{A}{9H_0^2 + \omega^2}\left( 3H_0 \sin\left(\omega t + \alpha\right)
\right. \nn
&& \left. - \omega \cos\left(\omega t + \alpha\right)\right)\ ,
\eea
with a constant $\rho_0$. This corresponds to de Sitter universe.

On the other hand, when  other contributions to the total energy
density can be neglected, as in the late-time  universe, by using
(\ref{BBB1}), one gets
\be
\label{B40b}
\frac{d^2 a^{3/2}}{dt^2} +\frac{3}{4}\kappa^2 A\sin\left(\omega t + \alpha\right)a^{3/2}=0\ .
\ee
By defining a new variable $s$
\be
\label{B40c}
s\equiv \omega t + \alpha + \frac{\pi}{2}\ ,
\ee
one obtains a kind of Mathieu equation:
\be
\label{B40d}
0=\frac{d^2 a^{3/2}}{ds^2} + \frac{3\kappa^2}{4\omega^2}\cos s\, a^{3/2}\ ,
\ee
whose solution is given by
\be
\label{B40e}
a^{3/2}=\sum_{n=0}^\infty c_n \cos(nt) + \sum_{n=1}^\infty s_n \sin (nt)\ .
\ee
Here the coefficients $c_n$ and $s_n$ are given by recursively solving
  the following equations:
\bea
\label{B40f}
&& c_0=c\ ,\quad c_1=0 \nn
&& -n^2 c_n + \frac{q}{2}\left(c_{n-1} + c_{n+1}\right)=0\ \left(n\geq 1\right)\ , \nn
&& s_1=s\ ,\quad s_2=- \frac{2}{q}s\ ,\nn
&& -n^2 s_n + \frac{q}{2}\left(s_{n-1} - s_{n+1}\right)=0\ \left(n\geq 2\right)\ ,\nn
&& q\equiv \frac{3\kappa^2}{4\omega^2}\ .
\eea
Hence, $a$ has a periodicity $1/\omega$. In the expression (\ref{B40e}),
$a$ is not always positive.
Then physically the regions where $a^{3/2}$ is not negative could be allowed
and the points $a=0$
could correspond to Big Bang/Big Crunch/Big Rip\cite{brett}.

We should note that the expressions of $\rho_0$ in (\ref{B37}) and (\ref{B40})
are not always
positive. Then only the period(s) where $\rho_0$ is positive could be allowed
in the real
universe.

\section{Coupling with the matter}

\subsection{No direct interaction between dark energy and matter}

Let us now include the matter. For simplicity, we consider the matter with
constant
EoS parameter $w_m$ so that
 the matter energy density $\rho_m$  is given by
\be
\label{B41}
\rho=\rho_0 \left(\frac{a(t)}{a(t_0)}\right)^{-3(1+w_m)}\ .
\ee
In case of (\ref{B11}), the total energy density is given by
\be
\label{B42}
\rho_{\rm tot} = \rho_c + x^{\sigma} \left[C_1 J_{\nu}(\hat a x^{\lambda}) +
C_2 J_{-\nu}(\hat a x^{\lambda})\right] + \rho_0 x^{-3(1+w_m)} \,,
\ee
and the Hubble rate $H$ is given by
\bea
\label{B43}
H&=&\kappa\left\{ \frac{1}{3}\left(\rho_c + x^{\sigma} \left[C_1 J_{\nu}(\hat a
x^{\lambda}) +
C_2 J_{-\nu}(\hat a x^{\lambda})\right] \right.\right. \nn
&& \left.\left. + \rho_0 x^{-3(1+w_m)}\right) \right\}^{1/2} \,.
\eea
In future, $x$  becomes large, then the Hubble rate $H$ goes to a constant
(with oscillations):
\be
\label{B44}
H\to \kappa\sqrt{\frac{\rho_0}{3}}\ ,
\ee
which tells $w_{\rm eff}\to -1$.
On the other hand, in the early universe, $x$ should be small. Hence, one
finds
\bea
\label{B45}
H&=&\kappa\left\{ \frac{1}{3}\left(\rho_c + x^{\sigma} \left[ C_1
\left(\frac{\hat a x^{\lambda}}{2}\right)^\nu +
C_2 \left(\frac{\hat a x^{\lambda}}{2}\right)^{-\nu}\right] \right. \right. \nn
&& \left.\left. + \rho_0 x^{-3(1+w_m)}\right) \right\}^{1/2} \ .
\eea
If $-3(1+w_m)<\sigma - \lambda$, the contribution from matter becomes
dominant and Hubble rate is
\be
\label{B46}
H=\kappa\left\{ \frac{\rho_0}{3} \right\}^{1/2}x^{-3(1+w_m)/2}\ ,
\ee
which gives, as well-known,
\be
\label{B47}
H\sim \frac{\frac{2}{3(1+w_m)}}{t}\ .
\ee
On the other hand
if $\sigma - \lambda \nu < -3(1+w_m)<0$, Hubble rate is
\be
\label{B48}
H=\kappa\left\{ \frac{C_2}{3}  \left(\frac{\hat a}{2}
\right)^{-\nu}\right\}^{1/2}
x^{\left(\sigma - \lambda\nu\right)/2} \ ,
\ee
which gives
\be
\label{B49}
H=\frac{-\frac{2}{\sigma - \lambda\nu}}{t}\ .
\ee
By comparing (\ref{B49}) with (\ref{B47}) or (\ref{B11}), it follows that
the
effective EoS parameter is given by
\be
\label{B50}
w_{\rm eff}= -1 - \frac{\sigma - \lambda \nu}{3}\ .
\ee

For the model in (\ref{B20}),  solving (\ref{B21}), we find the
$t$-dependence of $\rho$.
Then the FRW equation gives
\be
\label{B48b}
\frac{3}{\kappa^2}H^2 =\rho(t) + \rho_0
\left(\frac{a(t)}{a(t_0)}\right)^{-3(1+w_m)}\ .
\ee
For the case (\ref{B22}), when $t$ is large enough, the second term in the
r.h.s. of (\ref{B48b}) could be neglected and
we will obtain (\ref{B25}). If $c_0$ in (\ref{B22}) is small enough, when
$t\sim t_0$, the second term in (\ref{B48b})
could be dominant and we may obtain (\ref{B47}).
For the case (\ref{B26}), in the early universe, where $a$ is small, the
second term in (\ref{B48b})
could be dominant and one obtains (\ref{B47}), again.
Especially for the dust $w_m=0$, we find $H\sim \frac{2/3}{t}$,
that is, $a\sim t^{\frac{2}{3}}$.
In the late time
universe, the first term could be dominant
and one gets (\ref{B26}).

Three years WMAP data are recently analyzed in Ref.\cite{Spergel}, which
shows that
the combined analysis of WMAP with supernova Legacy
survey (SNLS) constrains the dark energy equation of state $w_{DE}$ pushing it
towards the cosmological constant. The marginalized best fit values of the
equation of state parameter at 68$\%$ confidance level
are given by $-1.14\leq w_{DE} \leq -0.93$. In case of a prior that universe is
flat, the combined data gives  $-1.06 \leq w_{DE} \leq -0.90 $.

In our models, as shown in (\ref{B12}), (\ref{B13}), (\ref{BBB4}), and
(\ref{B44}), the effective EoS parameter is $w_{\rm eff}\sim -1$ and
there is no contradiction with the above WMAP data.
We should also note that when mater is coupled, we find $w_{\rm eff}\sim w_m$
in the early universe, as in (\ref{B47}).
Thus when $w_m<-1/3$, there should occur the transition from deceleration
to
acceleration.

\subsection{ Dark energy interacting with matter}

Generally speaking, the matter interacts with the dark energy. In such a
case,
the total energy density  $\rho_{\rm tot}$ consists of
the contributions from the dark energy and the matter:
$\rho_{\rm tot}=\rho + \rho_m$. If we define, however, the matter energy
density $\rho_m$ properly, we can also {\it define} the matter pressure $p_m$
and the dark energy pressure $p$ by
\be
\label{Sep1}
p_m\equiv -\rho_m + \frac{\dot \rho}{3H}\ , \quad
P\equiv P_{\rm tot} - P_m\ .
\ee
Here $P_{\rm tot}$ is the total pressure. Hence, the  matter and dark energy
satisfy the energy conservation laws separately,
\be
\label{Sep2}
\dot \rho_m + 3H\left(\rho_m + P_m\right)=0\ ,\quad
\dot \rho + 3H\left(\rho + P\right)=0\ .
\ee
In case, however, that the EoS parameter $w_m$ for the matter
is almost constant, one may write the
conservation law as
\be
\label{B51}
\dot\rho_m + 3H\left(1+w_m\right)\rho_m = Q\ ,
\ee
and therefore for the dark energy
\be
\label{B52}
\dot\rho + 3H\left(\rho + P\right) = -Q\ ,
\ee
so that the total energy density and the pressure satisfy the conservation law:
\be
\label{B53}
\dot\rho_{\rm tot} + 3H\left(\rho_{\rm tot} +P_{\rm tot}\right)\rho_m = 0\ .
\ee
In (\ref{B51}), $Q$ expresses the shift from the constant EoS parameter case.

As an example, we consider the case that $Q$ is given by a function $q=q(a)$ as
\be
\label{B54}
Q=Ha q'(a) \rho_m\ .
\ee
Combining (\ref{B54}) with (\ref{B51}), one gets
\be
\label{B55}
\rho_m = \rho_{m0} a^{-3(1 + w_m)} \e^{q(a)}\ .
\ee
Here $\rho_{m0}$ is a constant of the integration.
Hence, the conservation law (\ref{B1}) is modified, through (\ref{B52}) as
\be
\label{B56}
\dot\rho + 3H\left(\rho + P\right) = -\rho_{m0} H a^{-\left(2+3w_m\right)}
q'(a) \e^{r(a)}\ ,
\ee
and (\ref{B6}) is also modified as
\be
\label{B57}
\frac{1}{3} x \frac{d \rho}{dx} + \rho = - P  - S(x) \ .
\ee
Here
\be
\label{B58}
S(x) \equiv
  - \frac{\rho_{m0}}{3}  \left(a(t_0)x\right)^{-\left(2+3w_m\right)}
q'\left(a(t_0)x\right)
\e^{q\left(a(t_0)x\right)} \ .
\ee
Note that Eq.(\ref{B8}) is also modified: it now contains the
inhomogeneous terms:
\bea
\label{B59}
&& x^2 \frac{d^2 \rho}{dx^2} + x \frac{d \rho}{dx} \left(4 + \frac{1}{\xi}
\right)
+ \frac{3}{\xi} [\rho + f(\rho,x)] \nn
&=& 3x\frac{dS(x)}{dx} + \frac{3}{\xi}S(x) \nn
&=& 3 x^{1 - 1/\xi} \frac{d}{dx}\left(x^{1/\xi} S(x)\right)\ .
\eea
Let a (special) solution of (\ref{B8}) be $\rho=\rho_s(x)$. Then in case of
(\ref{B9}) with
$\gamma(x) = \gamma_0 + \alpha x^m$, the general solution corresponding to
(\ref{B11}) is given by
\be
\label{B60}
\rho = \rho_s(x) + x^{\sigma} \left[C_1 J_{\nu}(\hat a x^{\lambda}) +
C_2 J_{-\nu}(\hat a x^{\lambda})\right] \,.
\ee
where $\rho_c$ should be included in $\rho_s(x)$.
It is also noted that the initial conditions are relevant to determine
$C_1$
and $C_2$ but irrelevant
for $\rho_s(x)$.
As an example, we can find
\be
\label{B61}
\rho_s(x) = \rho_c + \rho_0 x^\eta\ ,
\ee
with constants $\rho_0$ and $\eta$ when
\bea
\label{B62}
&& \e^{q(a)}= \e^{q_0} \nn
&& - \frac{\rho_0}{\rho_{m0}}\left[
\frac{\left(\eta^2 - 3\eta/4 + \eta/\xi +
3\gamma_0/\xi\right)a_0^{-\eta}a^{\eta + 3(1+w_m)} }{(\eta + 1/\xi)
\left(\eta + 3(1+w_m)\right)} \right. \nn
&& \left. + \frac{(3\alpha/\xi)a_0^{-m-\eta}a^{m+ \eta + 3(1+w_m)} }{(m+\eta +
1/\xi)\left( m+\eta +3(1+w_m) \right)} \right] \nn
&& - \frac{3s_0 a_0^{1/\xi}a^{ -1/\xi + 3 ( 1+w_m) }}{\rho_{m0}\left( -1/\xi +
3 ( 1+w_m) \right)}\ .
\eea
which gives
\bea
\label{B63}
S&=& \frac{\rho_0}{3}\left[ \frac{\left( \eta^2 - 3\eta/4 + \eta/\xi +
3\gamma_0/\xi \right)x^\eta}{\eta + 1/\xi} \right. \nn
&& \left. + \frac{3\alpha x^{m+\eta}}{\left(m + \eta + 1/\xi\right)\xi} \right]
+ s_0 x^{-1/\xi}\ .
\eea
In (\ref{B62}) and (\ref{B63}), $q_0$, $s_0$ are constants and $a_0\equiv
a(t_0)$.
In case that
\be
\label{B64}
\eta + 3(1+w_m),\ m+ \eta + 3(1+w_m),\ -1/\xi + 3 ( 1+w_m) <0\ ,
\ee
we find $\e^{q(a)}\to \e^{q_0}$ when $a$ becomes large, that is, in the
late
time universe.
Thus, $\rho_m \to \rho_{m0}\e^{q_0} a^{-(2+3w_m)}$.
Furthermore if $\eta<0$, we find $\rho_s\to \rho_0$, that is, $H$ goes to
a constant,
which may lead to the asymptotically deSitter space.
Clearly, for more complicated coupling $Q$, more sophisticated
accelerating cosmology may be constructed.

\section{Discussion}

In summary, we discussed the constrained EoS for cosmic fluid where
the relaxation equation for pressure is introduced. It is shown
that such EoS is equivalent to usual inhomogeneous EoS \cite{inh}
which contains scale factor dependent terms. Subsequently,
the generalized inhomogeneous EoS with time derivatives of pressure
 is presented.  For the number of explicit examples,
the accelerating dark energy cosmology as follows from
such EoS cosmic fluid is constructed. It turns out to be the
asymptotically de Sitter universe or oscillating universe with long
accelerating phase and transtion from deceleration to acceleration.
The consistent coupling of such constrained EoS dark fluid
with matter is discussed. It is shown that emerging FRW cosmology
may be consistent with three years WMAP data.

Of course, there are many ways to generalize the EoS
for cosmic fluid and to investigate the corresponding impact
of such generalization to dark cosmos. The physics behind such
generalization remains to be quite obscure
(as dark energy itself and its sudden appearence). At best,
this may be considered as some phenomenological approximation.
Nevertheless, having in mind, that most of modern attempts to
understand dark energy including strings/M-theory, brane-worlds,
 modified gravity, etc lead to effective description  in terms
 of cosmic fluid with unusual form of EoS, it turns out to
be extremely powerful approach.
 From another side, the reconstruction of the cosmic
fluid EoS may be done for any given cosmology compatible with
observational data which may finally select the true dark energy theory.

\section*{Acknoweledgements}

We are very grateful to A. Balakin for stimulating discussions and
participation at the early stage of this work.
The research by SDO was supported in part by LRSS project n4489.2006.02
(Russia), by RFBR grant 06-01-00609 (Russia), by project FIS2005-01181
(MEC, Spain) and by the project 2005SGR00790 (AGAUR,Catalunya, Spain)
and the
 research by S.N. was  supported in part by YITP computer facilities.

\end{document}